\documentclass[prb,twocolumn,showpacs,floatfix]{revtex4}
\usepackage{graphicx}
\usepackage{amssymb}
\usepackage{amsmath}
\usepackage{xspace}
\usepackage{url}

\let \Re \relax
\DeclareMathOperator{\Re}{Re}

\begin{document} 
\title{Optical signatures of the Charge of a Dielectric Particle in a Plasma}

\author{R. L. Heinisch, F. X. Bronold, and 
H. Fehske}
\affiliation{Institut f{\"ur} Physik,
             Ernst-Moritz-Arndt-Universit{\"a}t Greifswald,
             17487 Greifswald,
             Germany}

\date{\today}
\begin{abstract}
With an eye on dust particles immersed into an ionized gas, we study the effect of a negative charge on the scattering of light by a dielectric particle with a strong transverse optical phonon resonance in the dielectric constant. Surplus electrons alter the scattering behavior of the particle by their phonon limited conductivity in the surface layer (negative electron affinity) or in the bulk of the particle (positive electron affinity). We identify a charge-dependent increase of the extinction efficiency for low frequencies, a shift of the extinction resonance above the transverse optical phonon frequency, and a rapid variation of the polarization angles over this resonance. These effects could be used for non-invasive optical measurements 
of the charge of the particle.
\end{abstract}
\maketitle

\section{Introduction}

Charged dust particles embedded in a plasma environment are an ubiquitous phenomenon in nature.\cite{Mendis02,Ishihara07} 
They are found in the interstellar medium,\cite{Spitzer82,Mann08} planetary magnetospheres,\cite{GGM84} the upper 
atmosphere,\cite{FR09} and in industrial plasmas.\cite{Hollenstein00} Dusty laboratory plasmas,\cite{PM02} 
containing self-organized dust clouds, serve moreover as model systems for studying the dynamic 
behavior of strongly Coulomb-correlated systems of finite extent.

From the plasma physics point of view, the most important property of a dust particle is the charge it
accumulates from the plasma. It controls the coupling to other dust particles and to external electromagnetic 
fields as well as the overall charge balance of the plasma. As a consequence various methods have been devised to 
measure the particle charge. They range from force 
balance methods for particles drifting in the plasma~\cite{KRZ05} or trapped in the plasma sheath~\cite{TLA00,CJG11} 
to methods based on wave dispersion,\cite{HMP97} normal mode analysis,\cite{Melzer03} and dust cluster rotation.\cite{CGP10} 
Yet, the precise determination of the particle charge in a plasma environment remains a challenge. Methods independent of 
the plasma parameters,\cite{HMP97,Melzer03,CGP10} which are 
usually not precisely known, require specific experimental configurations, long measurement times or cannot yield the 
charge of individual dust particles. The phase-resolved resonance 
method,\cite{CJG11} for instance, allows only a precise relative charge measurement. For an absolute charge measurement 
the potential profile in the vicinity of the particle has to be additionally obtained by Langmuir probe measurements 
which however are only about 20\% accurate. Thus an optical measurement of the particle charge, independent of 
the plasma parameters, would be extremely useful.
 
The scattering of light by a small particle (Mie scattering~\cite{Mie08}) encodes---at least in principle---the particle 
charge.\cite{BH77,BH83,KK07,KK10,HCL10,HBF12b} It enters the scattering coefficients through the electrical conductivity of 
the surplus electrons which modifies either the  boundary conditions for the electromagnetic fields or the polarizability 
of the particle. To assess how and at which frequencies charges are revealed by the Mie signal requires however not only a 
microscopic calculation of the surface and bulk conductivities but also a detailed analysis of the conductivities'
impact on the different scattering regimes the particle's dielectric constant gives rise to. 

So far, the dependence of the Mie signal on the particle charge has not been investigated 
systematically. In our previous work~\cite{HBF12b} we made a first step to clarify this issue which has also been raised by Rosenberg.\cite{Rosenberg12} We 
identified the extinction at anomalous optical resonances of dielectric particles with a strong transverse optical 
(TO) phonon resonance in the dielectric constant to be sensitive to surplus electrons. In the present work we give a 
more comprehensive survey of Mie scattering by small negatively charged dielectric particles. Our aim is to identify 
over the whole frequency range, not only in the vicinity of anomalous resonances, features in the Mie signal which 
respond to surplus electrons. From these features the surplus electron density 
of the particle could be determined optically via light scattering. 

After a brief outline of the Mie theory of light scattering by small charged particles in the next section, we present
in Section III results for the four generic scattering features which occur for a charged dielectric particle with 
a strong resonance in the dielectric constant at the TO phonon frequency $\omega_{TO}$: low-frequency scattering, 
ordinary resonances below $\omega_{TO}$, anomalous resonances above $\omega_{TO}$, 
and high-frequency scattering. We investigate the intensity 
of the Mie signal and its polarization. Thereby we include ellipsometric techniques into our considerations. 
Section IV finally summarizes the results and points out possibilities for an
optical measurement of the particle charge.

\section{Theory}
\label{sec_theory}

The scattering behavior of an uncharged dielectric particle is determined by its radius $a$ and frequency-dependent dielectric 
constant $\epsilon(\omega)$. For a negatively charged dielectric particle light scattering is also influenced by the electric conductivity of the surplus electrons. Whether surplus electrons are trapped inside the particle or in a surface layer around it depends on the electron affinity $\chi$ of the particle.\cite{HBF12b} 

For $\chi<0$, as it is the case for instance for MgO and LiF,\cite{RWK03}  the conduction band inside the dielectric is above the potential outside the particle. Electrons do not penetrate into the dielectric. Instead they are trapped in the image potential in front of the surface.\cite{HBF12,HBF11} The image potential is due to a surface mode associated with the TO phonon. The interaction of an electron with the surface mode  comprises a static part, which induces the image potential,\cite{EM73,Barton81} and a dynamic part, which enables momentum relaxation parallel to the surface limiting the surface conductivity.\cite{KI95} The phonon-limited surface conductivity $\sigma_s$, calculated in our previous work~\cite{HBF12b} using the memory function approach,\cite{GW72} 
modifies the boundary condition for the magnetic field at the surface of the grain.\cite{BH77}

For $\chi>0$, as it is the case for instance for Al$_2$O$_3$, Cu$_2$O and PbS, the conduction band inside the dielectric lies below the potential outside the particle. Electrons thus accumulate in the conduction band where they form an extended space charge.\cite{HBF12} Its width, limited by the screening in the dielectric, is typically larger than a micron. For micron-sized particles we can thus  assume a homogeneous electron distribution in the bulk. The bulk conductivity is limited by scattering with a longitudinal optical (LO) phonon~\cite{Mahan90} and can be also calculated~\cite{HBF12b} within the memory function approach. 
The bulk conductivity of the surplus electrons $\sigma_b$ leads to an additional polarizability per volume $\alpha=4\pi i \sigma_b / \omega$ which alters the refractive index.

The scattering behavior of the particle is controlled by the scattering coefficients. They are determined by expanding the incident ($i$) plane wave into spherical vector harmonics and matching the reflected ($r$) and transmitted ($t$) waves at the boundary of the sphere.\cite{Stratton41,BH83}  In the case of $\chi>0$ the boundary conditions at the surface are given by $\mathbf{\hat{e}}_r \times (\mathbf{C}_i+ \mathbf{C}_r-\mathbf{C}_t)=0$ for $\mathbf{C}=\mathbf{E},\mathbf{H}$.  For $\chi<0$ the surface charges may sustain a surface current $\mathbf{K}=\sigma_s \mathbf{E}_\parallel$ which is induced by the parallel component of the electric field and proportional to the surface conductivity. This changes the boundary condition for the magnetic field to
$\mathbf{\hat{e}}_r\times (\mathbf{H}_i+\mathbf{H}_r-\mathbf{H}_t)=4\pi\mathbf{K}/c$,
 where $c$ is the velocity of light. The boundary condition for the electric field is still $\mathbf{\hat{e}}_r \times (\mathbf{E}_i+ \mathbf{E}_r-\mathbf{E}_t)=0$. The refractive index of the particle  $N=\sqrt{\epsilon}$ ($\chi<0$) or $N=\sqrt{\epsilon+\alpha}$ ($\chi>0$). Matching the fields at the particle surface gives the scattering coefficients~\cite{BH77}
\begin{equation}
a_n^r=-\frac{F_{n}^a}{F_{n}^a+iG_{n}^a}, \quad b_n^r=-\frac{F_{n}^b}{F_{n}^b+iG_{n}^b},
\end{equation}
where
\begin{align}
F_{n}^a&=\psi_n(N\rho) \psi^\prime_n(\rho)-\left[N\psi_n^\prime(N\rho) -i\tau \psi_n(N\rho)\right] \psi_n(\rho), \\
G_{n}^a&=\psi_n(N\rho) \chi^\prime_n(\rho)-\left[N\psi_n^\prime(N \rho ) -i\tau \psi_n(N\rho) \right] \chi_n(\rho), \\
F_{n}^b&=\psi_n^\prime(N\rho) \psi_n(\rho)-\left[N\psi_n(N\rho) +i\tau \psi_n^\prime(N\rho)\right] \psi_n^\prime(\rho), \\
G_{n}^b&=\psi_n^\prime(N\rho) \chi_n(\rho)- \left[ N\psi_n(N\rho) +i\tau \psi_n^\prime(N\rho)\right] \chi_n^\prime(\rho)
\end{align}
with the dimensionless surface conductivity $\tau(\omega)=4\pi \sigma_s(\omega)/c$  ($\chi<0$) or $\tau=0$  ($\chi>0$). The size parameter $\rho=ka=2\pi a /\lambda$ where $\lambda$ is the wavelength and $\psi_n(\rho)=\sqrt{\pi \rho/2}J_{n+1/2}(\rho)$, $\chi_n(\rho)=\sqrt{\pi \rho/2}Y_{n+1/2}(\rho)$ with $J_n$ the Bessel and $Y_n$ the Neumann function.
The efficiencies for extinction ($t$) and scattering ($s$) are 
\begin{align}
Q_t&=-\frac{2}{\rho^2}\sum_{n=1}^\infty (2n+1) Re(a_n^r+b_n^r) \\
Q_s&=\frac{2}{\rho^2 }\sum_{n=1}^\infty (2n+1) (|a_n^r|^2+|b_n^r|^2)
\end{align}
from which the absorption efficiency $Q_a=Q_t-Q_s$ can be also obtained.

An important special case is the scattering by small particles, for which $\rho \ll 1$. Inspired by the expressions used in  
Ref.~\onlinecite{LTT07} we write in this case $F_n^a=N^{n+1}f_n^a/(2n+1)$ and $G_n^a=N^{n+1}g_n^a/(2n+1)$ with
\begin{widetext}
\begin{align}
f_n^a=&\frac{2^{2n}(n+1)!n!}{(2n+1)!(2n)!}\rho^{2n+1}\left(\frac{i\tau}{n+1}\rho+\frac{N^2-1}{(n+1)(2n+3)}\rho^2
+\mathcal{O}(\rho^3)  \right),
\label{fa}\\
g_n^a=&2n+1-i\tau\rho+\frac{1-N^2}{2}\rho^2 +\mathcal{O}(\rho^3), \label{ga}
\end{align}
and similarly $F_n^b=N^n f_n^b/(2n+1)$ and $G_n^b=N^n g_n^b/(2n+1)$ with
\begin{align}
f_n^b=& \frac{2^{2n} n!(n+1)!}{(2n)!(2n+1)!}\rho^{2n+1}\left(1-N^2-i(n+1)\frac{\tau}{\rho}+\mathcal{O}(\rho)\right), \label{fb} \\
g_n^b=&-(n+1)-nN^2-in(n+1)\frac{\tau}{\rho}\nonumber \label{gb}+\left[\frac{(n+3)N^2+nN^4}{2(2n+3)}-\frac{(n+1)+(n-2)N^2}{2(2n-1)}\right]\rho^2 \nonumber \\
&+\left[-i \frac{(n+1)(n-2)}{2(2n-1)}+i\frac{n(n+3)}{2(2n+3)}N^2\right] \tau \rho +\mathcal{O}(\rho^3)~.
\end{align}
\end{widetext}

The leading scattering coefficients for small uncharged particles are $b_1\sim \mathcal{O}(\rho^3)$ and 
$a_1, b_2\sim \mathcal{O}(\rho^5)$. For them
\begin{align}
f_1^a&=i\frac{\tau}{3}\rho^4+\frac{N^2-1}{15}\rho^5, \quad g_1^a=3-i\tau \rho+\frac{1-N^2}{2}\rho^2, \label{fg_a1}
\end{align}
\begin{align}
f_1^b&=-i\frac{4\tau}{3}\rho^2+\frac{2(1-N^2)}{3}\rho^3, \quad g_1^b=-2-N^2-i2\frac{\tau}{\rho} \label{fg_b1}, \\
f_2^b&=-i\frac{\tau}{5}\rho^4+\frac{1-N^2}{15}\rho^5, \quad g_2^b=-3-2N^2-i6\frac{\tau}{\rho} . \label{fg_b2}
\end{align}
Keeping only the coefficient $b_1$ the extinction efficiency $Q_t=-6\Re(b_1^r)/\rho^2$. Approximating $b_1^r=f/ig$ (we have neglected $f\sim \rho^3$ compared to $g\sim \rho^0$ in the denominator)  we obtain for the extinction efficiency 
\begin{align}
Q_t=\frac{12\rho \left(\epsilon^{\prime \prime}+\alpha^{\prime \prime}+2\tau^\prime/\rho \right)}{\left(\epsilon^{\prime }+\alpha^\prime+2-2\tau^{\prime \prime}/\rho \right)^2+\left(\epsilon^{\prime \prime}+\alpha^{\prime \prime}+2\tau^\prime/\rho \right)^2} \label{smallrhoext}
\end{align}
which is valid for small particles, that is, for $\rho\ll1$.

\section{Results}
\label{sec_results}

\begin{figure*}[t]
\includegraphics[width=\linewidth]{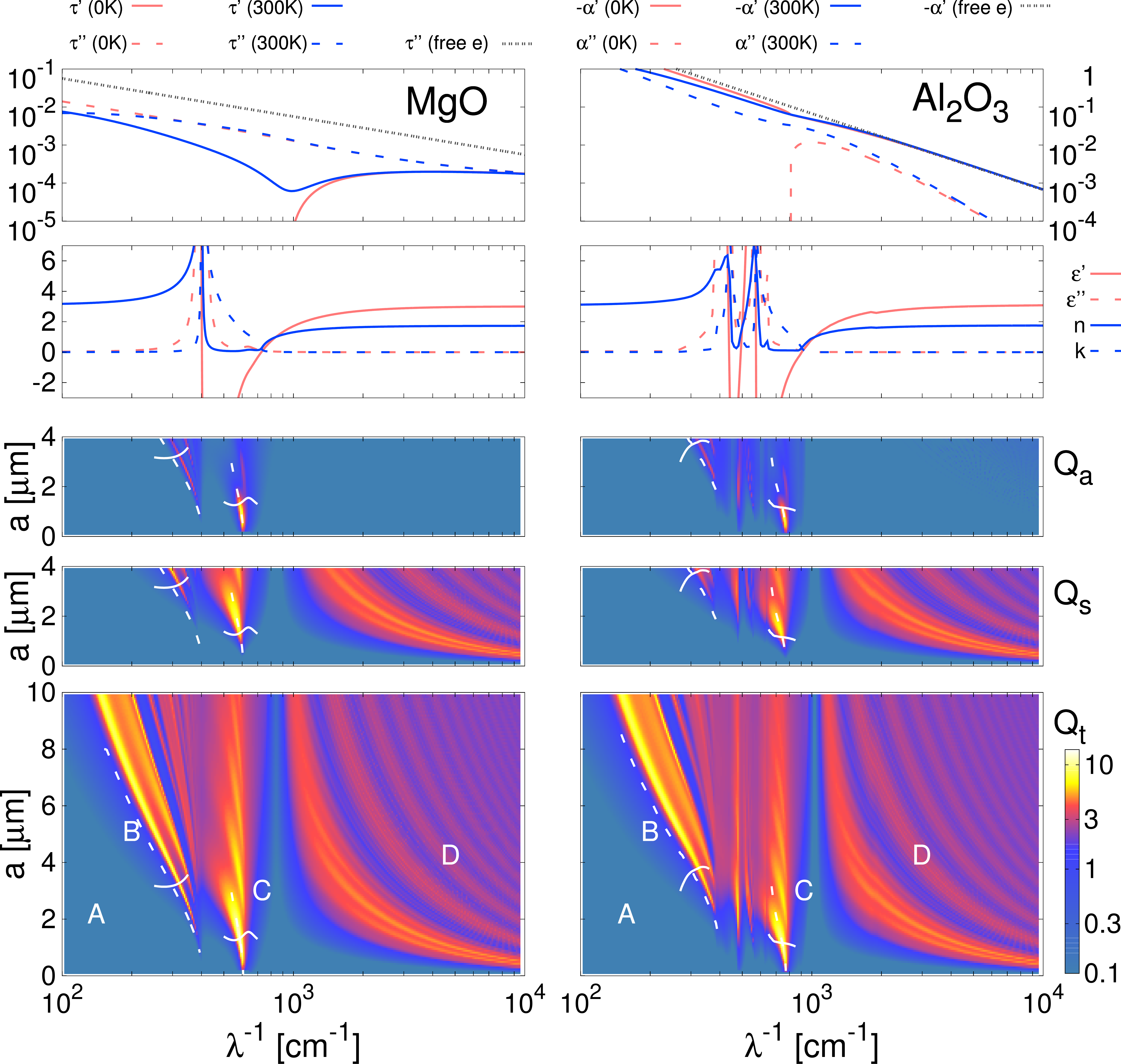}
\caption{(Color online) First row: Dimensionless surface conductivity $\tau=\tau^\prime+i\tau^{\prime \prime}$ for MgO for $n_s=10^{13}$cm$^{-2}$ (left) and polarizability of excess electrons $\alpha=\alpha^\prime+i\alpha^{\prime \prime}$ for Al$_2$O$_3$ for $n_b=3\times10^{17}$cm$^{-3}$ (right) as a function of the inverse wavelength $\lambda^{-1}$. Second row: Dielectric constant $\epsilon=\epsilon^\prime+i\epsilon^{\prime \prime}$ and refractive index $N=n+ik$ as a function of $\lambda^{-1}$. Third to fifth row: Absorption efficiency $Q_a$ (third row), scattering efficiency $Q_s$ (fourth row), and extinction efficiency \(Q_t\) (fifth row) as a function of $\lambda^{-1}$ and the particle radius $a$ for an uncharged MgO and Al$_2$O$_3$ particle. The labels indicate the four characteristic scattering regimes: low frequencies (A), ordinary resonances (B), anomalous resonances (C), and interference and ripple structure (D). The dashed lines give the approximate position of the $a_1^r$ (B) and the $b_1^r$  (C) resonance. The full lines give the approximate cross-over from absorption to scattering dominance of the resonances. }
\label{fig_over}
\end{figure*}

In the following we will discuss light scattering for a MgO ($\chi<0$) and an Al$_2$O$_3$ ($\chi>0$) particle (for material parameters see Ref.~\onlinecite{matpar}). The particle charge affects light scattering through the dimensionless surface conductivity $\tau=\tau^\prime+i\tau^{\prime \prime}$ (MgO) or the surplus electron polarizability $\alpha=\alpha^\prime+i\alpha^{\prime \prime}$ (Al$_2$O$_3$). Both $\tau$ and $\alpha$ are shown as a function of the inverse wavelength $\lambda^{-1}$ in the first row of 
Fig. \ref{fig_over}. They are small even for a highly charged particle with $n_s=10^{13}$ cm$^{-2}$ which corresponds to $n_b=3\times10^{17}$ cm$^{-3}$ for $a=1\mu$m.
For $T=300$~K $\tau^{\prime \prime} > \tau^\prime$ and $-\alpha^\prime > \alpha^{\prime \prime}$ except at very low frequencies. For $\lambda^{-1} \rightarrow 0$ the conductivities $\sigma_s$ and $\sigma_b$ tend to a real value so that $\tau^\prime > \tau^{\prime \prime}$ and $\alpha^{\prime \prime}>- \alpha^\prime$ for very small $\lambda^{-1}$.
 Both $\tau$ and $\alpha$ decrease with increasing $\lambda^{-1}$ and vary smoothly over the considered frequencies. Shown for comparison are also $\tau$ and $\alpha$ for $T=0$ where $\tau^\prime=0$ for $\lambda^{-1}<\lambda_s^{-1}=909$ cm$^{-1}$, the 
inverse surface phonon wavelength ($\alpha^{\prime \prime}=0$ for $\lambda^{-1}<\lambda_{LO}^{-1}=807$ cm$^{-1}$,
 the inverse LO phonon wavelength), since light absorption is possible only above $\lambda_s^{-1}$ (or $\lambda_{LO}^{-1}$).

The scattering behavior of the uncharged particles is primarily determined by the dielectric constants $\epsilon(\omega)$ 
(second row of Fig. \ref{fig_over}). For MgO it is dominated by a TO phonon at $\lambda^{-1}=401$ cm$^{-1}$. For Al$_2$O$_3$ 
two TO phonon modes at $\lambda^{-1}=434$ cm$^{-1}$ and $\lambda^{-1}=573$ cm$^{-1}$ dominate $\epsilon(\omega)$.
At frequencies well below the TO phonon resonance the dielectric constant tends towards its real static value $\epsilon_0$. 
In this regime (marker A in Fig. \ref{fig_over}) $\epsilon^{\prime\prime} \ll \epsilon^{ \prime}$. For constant radius $a$, the extinction efficiency $Q_t \rightarrow 0$ for $\lambda^{-1}\rightarrow 0$. 
Just below the TO phonon resonance (for Al$_2$O$_3$ below the lower TO-phonon)  $\epsilon^\prime$ is large and positive and 
$\epsilon^{\prime \prime} \ll \epsilon^{\prime}$ (except in the immediate vicinity of the resonance). This gives rise to 
ordinary optical resonances (marker B in Fig.~\ref{fig_over}).\cite{vandeHulst57}
Above the TO phonon resonance (for Al$_2$O$_3$ above the higher TO-phonon) $\epsilon^\prime <0$ and 
$\epsilon^{\prime \prime} \ll 1$. This entails anomalous optical resonances (marker C in 
Fig.\ref{fig_over}).\cite{FK68,TL06,Tribelsky11}
Far above the TO phonon resonance $\epsilon^\prime$ takes a small positive value and $\epsilon^{\prime \prime} \ll 1$. This 
gives rise to an interference and ripple structure (marker D in Fig. \ref{fig_over}).\cite{BH83}

In the context of plasma physics dielectric particles with a strong TO phonon resonance have already been studied theoretically as wave-length selective infra-red absorbers.\cite{Rosenberg12}
In the following we explore the modification of the  features  A--D  by surplus electrons. We are particularly interested in 
identifying dependencies in the optical signal which can be used as a charge diagnostic. 

\subsection{Low-Frequency Scattering}

\begin{figure}
\includegraphics[width=\linewidth]{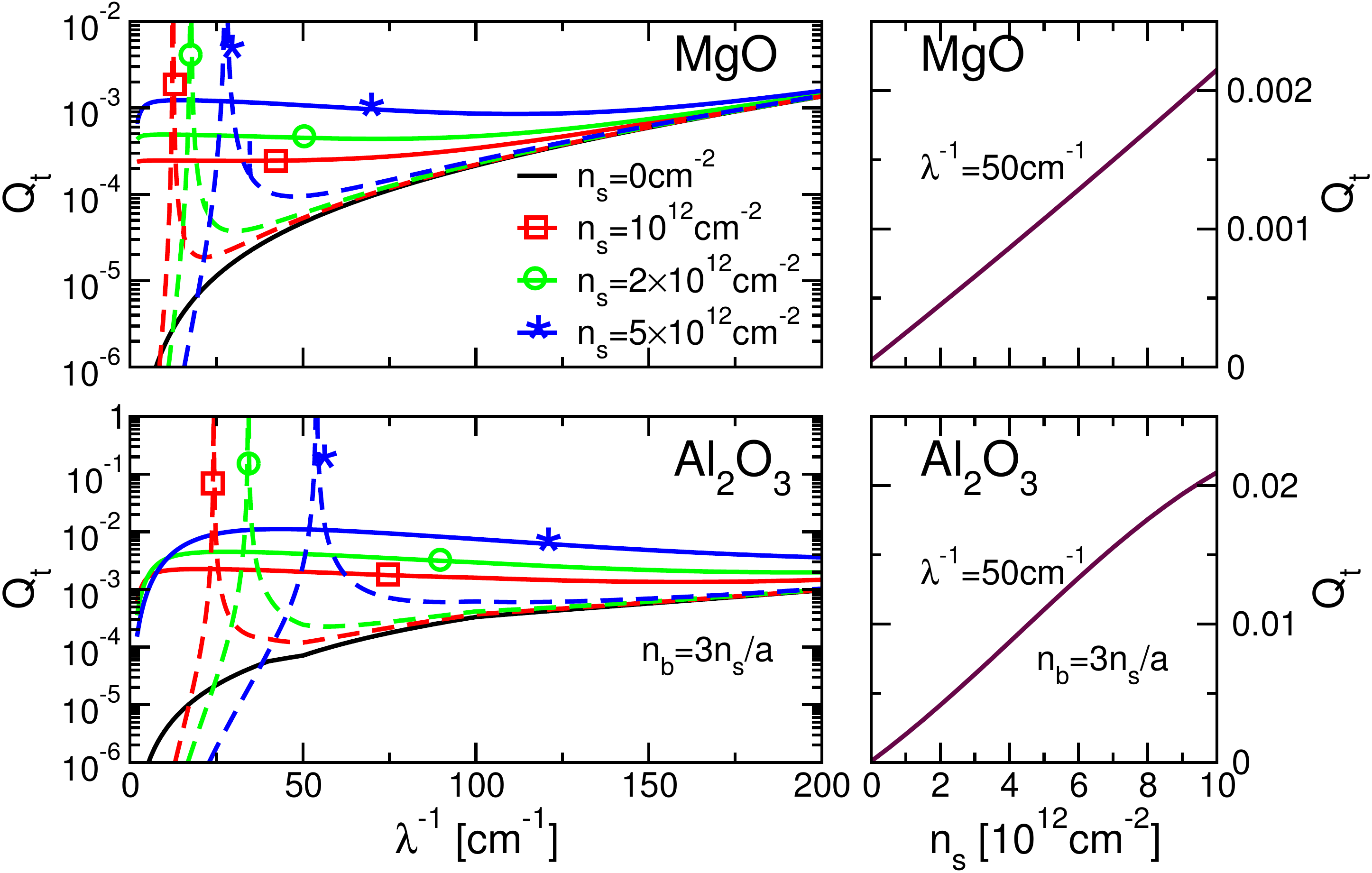}
\caption{(Color online) Left: Extinction efficiency \(Q_t\)  as a function of the inverse wavelength $\lambda^{-1}$  for a charged MgO particle (top) and a charged Al$_2$O$_3$ particle (bottom) with radius $a=1\mu$m at $T=300$ K. Full lines give the results for the phonon-limited surface or bulk conductivity, dashed lines show for comparison the results for free surface or bulk electrons. Right: Extinction efficiency as a function of the surface electron density for an MgO particle (or corresponding bulk electron density for Al$_2$O$_3$) for $\lambda^{-1}=50$ cm$^{-1}$.}
\label{fig_lowfreq}
\end{figure}

In the low frequency limit of scattering (marker A in Fig. \ref{fig_over}) the extinction efficiency $Q_t$ is relatively small. For $\lambda^{-1}<200$ cm$^{-1}$ particles with a radius of a few microns are small compared to the wavelength. In this limit the dominant scattering coefficient is $b_1^r$ and the extinction efficiency is given approximately by Eq. (\ref{smallrhoext}). Extinction is due to absorption which is controlled by $\epsilon^{\prime \prime}$. As $\epsilon^{\prime \prime}$ is small in this frequency range energy dissipation on the grain and thus extinction is inhibited. For  $\lambda^{-1}\rightarrow 0$, $\epsilon^{\prime \prime} \rightarrow 0$ and hence $Q_t \rightarrow 0$.

For charged dielectric particles light absorption is controlled not only by $\epsilon^{\prime \prime}$ but also by $\tau^\prime$ for $\chi<0$ and by  $\alpha^{\prime \prime}$ for $\chi>0$ which stem from the real part of the surface or bulk conductivity of the surplus electrons, respectively. For low frequency $\tau^\prime$ and $\alpha^{\prime \prime}$ are  larger than for higher frequencies and for $\lambda^{-1} \rightarrow 0$ even outweigh $\tau^{\prime \prime}$ and $-\alpha^{\prime} $ as the real parts of the surface and bulk conductivities tend to finite values whereas the imaginary parts vanish for $\lambda^{-1}=0$. This allows increased energy dissipation on charged dust particles entailing increased light absorption. Figure \ref{fig_lowfreq} shows this saturation of the extinction efficiency for charged particles. 

For comparison, we also show the results for free surface (MgO) or bulk electrons (Al$_2$O$_3$). In this case the conductivities are purely imaginary and the saturation of the extinction efficiency is not observed. Instead we find a plasmon resonance of the electron gas around or inside the particle. The resonance is located where $\Re(g_1^b)=0$ (with $g_1^b$ given by Eq. (\ref{fg_b1})). This discrepancy with results from the phonon-limited conductivities shows that in the low-frequency limit the model of free surplus electrons cannot offer even a qualitative explanation.

The saturation of the extinction efficiency for low frequencies could be employed as a charge measurement. Performing an extinction measurement at fixed wavelength would give an approximately linear increase of $Q_t$ with the surface density or bulk density of surplus electrons (see right panels of Fig. \ref{fig_lowfreq}).

\subsection{Ordinary Resonances}

Below the TO phonon resonance at $\lambda_{TO}^{-1}$ in the dielectric constant $\epsilon^\prime$ is large while $\epsilon^{\prime \prime}$ is still comparatively small (except at $\lambda_{TO}^{-1}$). The large positive $\epsilon^\prime$ (which entails a large positive real part of refractive index $N$) allows for ordinary optical resonances,\cite{vandeHulst57} which are clearly seen in Fig. \ref{fig_over}. The lowest resonance is due to the $a_1$ mode. The contribution of this mode to the extinction efficiency is $Q_{a_1}^t=-6 \Re (a_1^r)/\rho^2$. More generally, the 
  extinction efficiency due to one mode only reads $Q_{a,b_n}^t=2(2n+1) q_{a,b_n}^t / \rho^2$ where
\begin{equation}
q_{a,b_n}^t= \frac{f^{\prime } (f^\prime - g^{\prime \prime}) }{(f^\prime - g^{\prime \prime})^2+g^{\prime 2}} \label{rescon_ordinary}
\end{equation}
with $f=f^\prime+if^{\prime \prime}$ and $g=g^\prime+ig^{\prime \prime}$ (given for $\rho\ll1$ by Eqs. (\ref{fa})-(\ref{gb})). Note that we have neglected $f^{\prime \prime}$ as $\epsilon^{\prime \prime} \ll1$. The resonance is approximately located where $g^\prime=0$. This gives for $n=1$ the condition
\begin{align}
3+\tau^{\prime \prime}\rho +(1-\epsilon^\prime-\alpha^\prime)\rho^2/2=0. \label{rescon_a1}
\end{align}
The approximate resonance location for an uncharged sphere, obtained from $3 +(1-\epsilon^\prime)\rho^2/2=0$ is shown in Fig. \ref{fig_over} by the dashed line. It deviates somewhat from the true resonance location but captures its size dependence qualitatively. The contribution of one mode to absorption and scattering, respectively, is  
$Q_{a,b_n}^{a,s}=2(2n+1) q_{a,b_n}^{a,s} / \rho^2$ with
\begin{equation}
q_{a,b_n}^a= \frac{-f^\prime g^{\prime \prime}}{(f^\prime - g^{\prime \prime})^2+g^{\prime 2}}, \quad
q_{a,b_n}^s= \frac{f^{\prime 2}}{(f^\prime - g^{\prime \prime})^2+g^{\prime 2}} . \label{qs_qa_comp}
\end{equation}
For $f^\prime>-g^{\prime \prime}$ scattering outweighs absorption while absorption outweighs scattering for $-g^{\prime \prime}>f^{\prime}$. The boundary between the two regimes is given by $-g^{\prime\prime}=f^\prime$. For $n=1$ this gives for an uncharged particle
\begin{align}
\rho=\left(\frac{15}{2} \frac{\epsilon^{\prime \prime}}{\epsilon^\prime-1} \right)^\frac{1}{3} ,
\end{align}
which is shown in Fig. \ref{fig_over} by the solid line and agrees with the underlying contour.

\begin{figure}
\includegraphics[width=\linewidth]{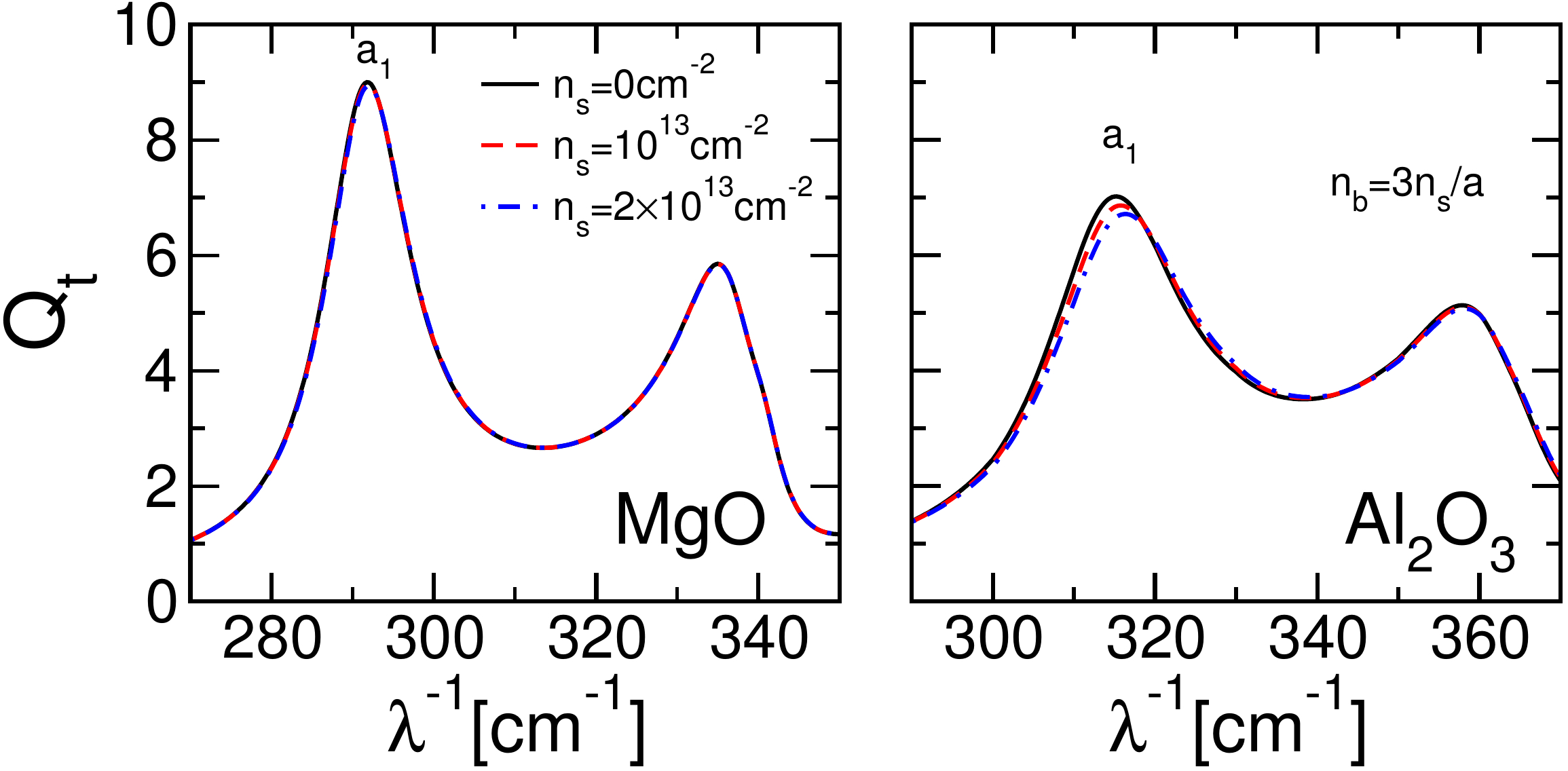}
\caption{(Color online) Extinction efficiency \(Q_t\)
as a function of the inverse wavelength for MgO (left) and Al$_2$O$_3$ (right) particles with radius \(a=4\mu\)m for \(n_s=0\),  \( 10^{13} \),  and $2\times 10^{13}$ cm$^{-2}$ (or corresponding bulk electron density $n_b=3n_s/a$) at \(T=300K\).}
\label{fig_ordres}
\end{figure}

Fig. \ref{fig_ordres} shows that the $a_1$ resonance is not shifted significantly by surplus charges. As the  charge enters $\sim \tau \rho$ or $\sim \alpha \rho^2$ the shift  cannot be increased by reducing the particle size. Ordinary resonances thus offer no possibility to measure the particle charge.

\subsection{Ripple and Interference Structure}

\begin{figure}
\includegraphics[width=\linewidth]{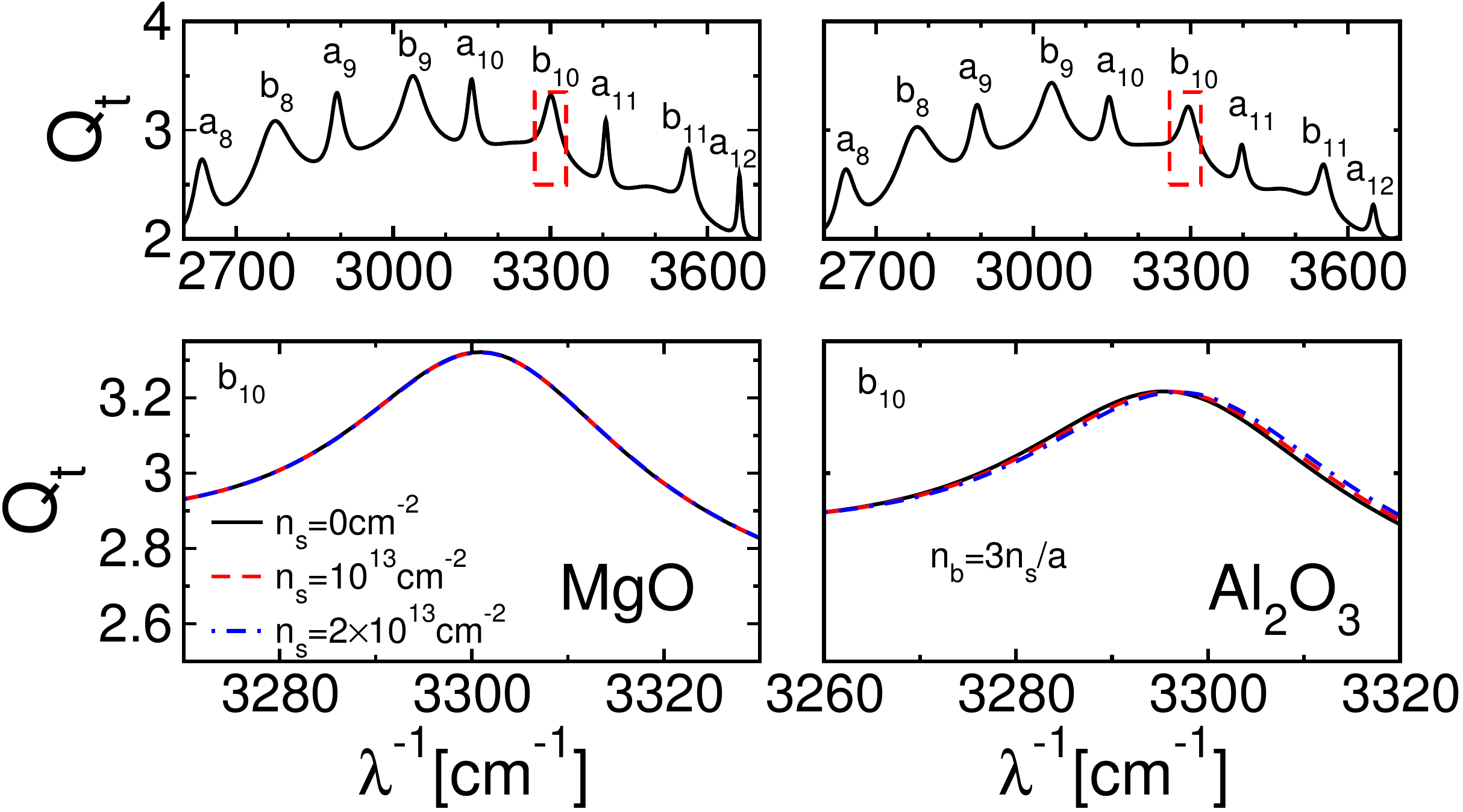}
\caption{(Color online) Top panel: Overview of the ripple and interference structure. Bottom panel: Extinction efficiency \(Q_t\) close to the $b_{10}$ ripple as a function of the inverse wavelength for MgO (left) and Al$_2$O$_3$ (right) particles with radius \(a=4\mu\)m for \(n_s=0\),  \( 10^{13} \)  and $2\times 10^{13}$ cm$^{-2}$ (or corresponding bulk electron density $n_b=3n_s/a$) at $T=300$ K.}
\label{fig_ri}
\end{figure}

Far above the highest TO phonon frequency (that is, for MgO and Al$_2$O$_3$ for $\lambda^{-1}>1000$ cm$^{-1}$) the extinction efficiency shows the typical interference and ripple structure of Mie scattering (marker D in Fig. \ref{fig_over}).\cite{BH83} It consists of a broad interference pattern superseded by fine ripples which are due to individual modes. They become sharper for larger frequencies. Figure \ref{fig_ri} shows the overall interference and ripple structure (top) and exemplifies the charge 
sensitivity of the ripple due to the mode $b_{10}$ (bottom). It is shifted only very slightly with increasing particle charge. This is due to the small values of the surface conductivity or the polarizability of the surplus electrons for $\lambda^{-1}>1000$ cm$^{-1}$. Thus the ripple structure is not a suitable candidate for a charge measurement either.

\subsection{Anomalous Resonances}

At the TO phonon resonance the real part of the dielectric constant changes sign. For $\lambda^{-1} > \lambda_{TO}^{-1}$  $\epsilon^\prime <0$ and $\epsilon^{\prime \prime}\ll1$. This gives rise to a series of anomalous optical resonances, which can be seen in Fig. \ref{fig_over} (marker C). They correspond to the resonant excitation of transverse surface modes of the sphere.\cite{FK68} For a metal particle they are tied to the plasmon resonance~\cite{TL06,Tribelsky11} whereas for a dielectric particle they are due to the TO-phonon. The resonances are associated with the  scattering coefficients $b_n$. The lowest resonance is due to the mode $b_1$. The resonance location is approximately given by $\Re(g_1^b)=0$, which according to Eq. (\ref{gb}) gives for an uncharged sphere
\begin{align}
-2-\epsilon^\prime +\left(-1-\frac{\epsilon^\prime}{10}+\frac{\epsilon^{\prime 2}-\epsilon^{\prime \prime 2}}{10}\right)\rho^2=0 .
\end{align}
This approximation, shown by the dashed line near marker C in Fig. \ref{fig_over}, agrees well with the underlying Mie contour. 

The higher resonances are scattering dominated, while the lowest resonance shows a cross-over from absorption to scattering dominance (see Fig. \ref{fig_over}). This cross-over can be understood from the contribution of the $b_1$ mode to the scattering and absorption efficiencies (given by Eq. (\ref{qs_qa_comp})). Absorption dominates for $-g^{\prime \prime}>f^\prime$, while scattering dominates for  $-g^{\prime \prime}<f^\prime$. The boundary between the two regimes lies where $-g^{\prime \prime}=f^\prime$. For $n=1$ this gives
\begin{equation}
\rho=\left( \frac{3}{2} \frac{\epsilon^{\prime \prime}}{1-\epsilon^\prime}\right)^\frac{1}{3}
\end{equation}
which agrees well with the Mie contour (see Fig \ref{fig_over}).

\begin{figure}[t]
\includegraphics[width=\linewidth]{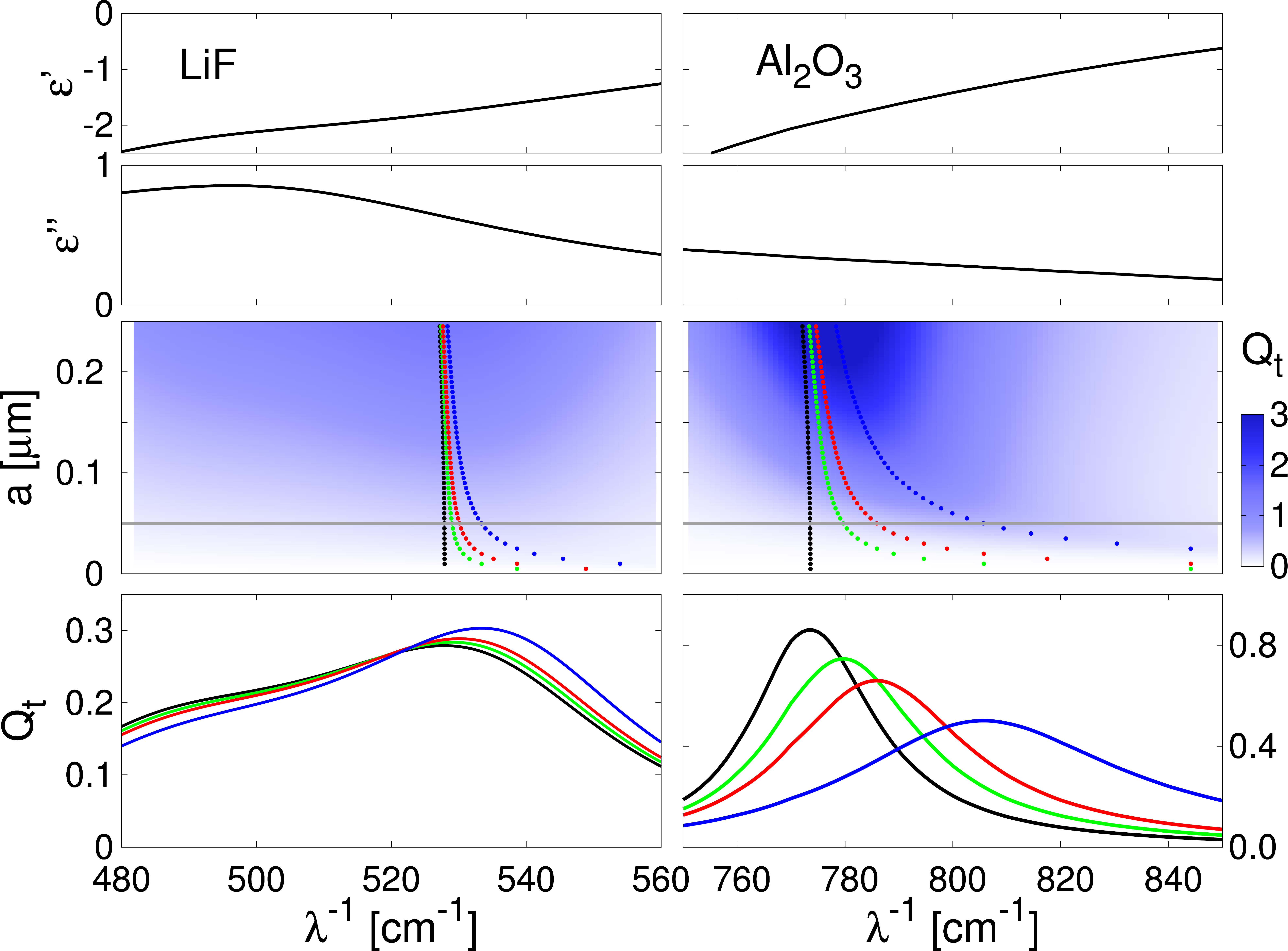}
\caption{(Color online)  Top panels: Real part $\epsilon^\prime$ and imaginary part $\epsilon^{\prime \prime}$ of the dielectric constant as a function of the inverse wavelength $\lambda^{-1}$. The maximum of $\epsilon^{\prime \prime}$ for LiF stems from a TO phonon mode at $503$ cm$^{-1}$. 
 Middle panel: Extinction efficiency $Q_t$ as a function of $\lambda^{-1}$ and the
radius $a$ for a LiF particle with $n_s= 5  \times 10^{12}$ cm$^{-2}$ (left) and an Al$_2$O$_3$ particle with $n_b = 3n_s /a$ (right) for
T = 300 K. The dotted lines indicate the extinction maximum for (from left to right) $n_s = 0$ (black), $10^{12}$ (green), $2 \times
10^{12}$ (red), and $5\times 10^{12}$ cm$^{-2}$ (blue). Bottom panel: Extinction efficiency $Q_t$ as a function of $\lambda^{-1}$ for different
surface electron densities (corresponding to the middle panel) and fixed radius  $ a = 0.05\mu$m. The extinction maximum is shifted to higher frequencies with increasing electron density.
}
\label{fig_anres}
\end{figure}

The anomalous resonances are sensitive to  small changes in $\epsilon$ and $\tau$ or $\alpha$. Surplus electrons lead to a blue-shift of the resonances.\cite{HBF12b} This effect is strongest for small particles with radius $a<1\mu$m. In the small particle limit the extinction efficiency is approximately given by Eq. (\ref{smallrhoext}). The resonance is located at
\begin{align}
\epsilon^\prime+\alpha^\prime+2-2\tau^{\prime \prime}/\rho=0 . \label{rescon_b1}
\end{align}
Compared to the resonance condition for ordinary resonances, Eq. (\ref{rescon_a1}), the charge sensitivity increases for small particles as surplus charges enter by $-2\tau^{\prime \prime}/\rho \sim n_s /a$ (for $\chi<0$) or $\alpha^\prime \sim n_b$ (for $\chi>0$). This shows that the resonance shift by the surplus electrons is primarily an electron density effect on the polarizability of the dust particle.\cite{HBF12b}

Figure \ref{fig_anres} shows the resonance shift for charged sub-micron-sized LiF~\cite{matpar} and Al$_2$O$_3$ particles. For Al$_2$O$_3$ the resonance shift is relatively large and the extinction resonance has a Lorentzian shape. As $\epsilon^{\prime}$ is well approximated linearly close to $-2$ and $\epsilon^{\prime \prime}$ varies only slightly this follows form Eq. (\ref{smallrhoext}). For LiF the shift is smaller and the lineshape is not Lorentzian. The reason is the minor TO phonon at  $\lambda^{-1}=503$cm$^{-1}$. This leads to a bump in $\epsilon^{\prime \prime}$ disturbing the Lorentzian shape.

\begin{figure*}[t]
\includegraphics[width=\linewidth]{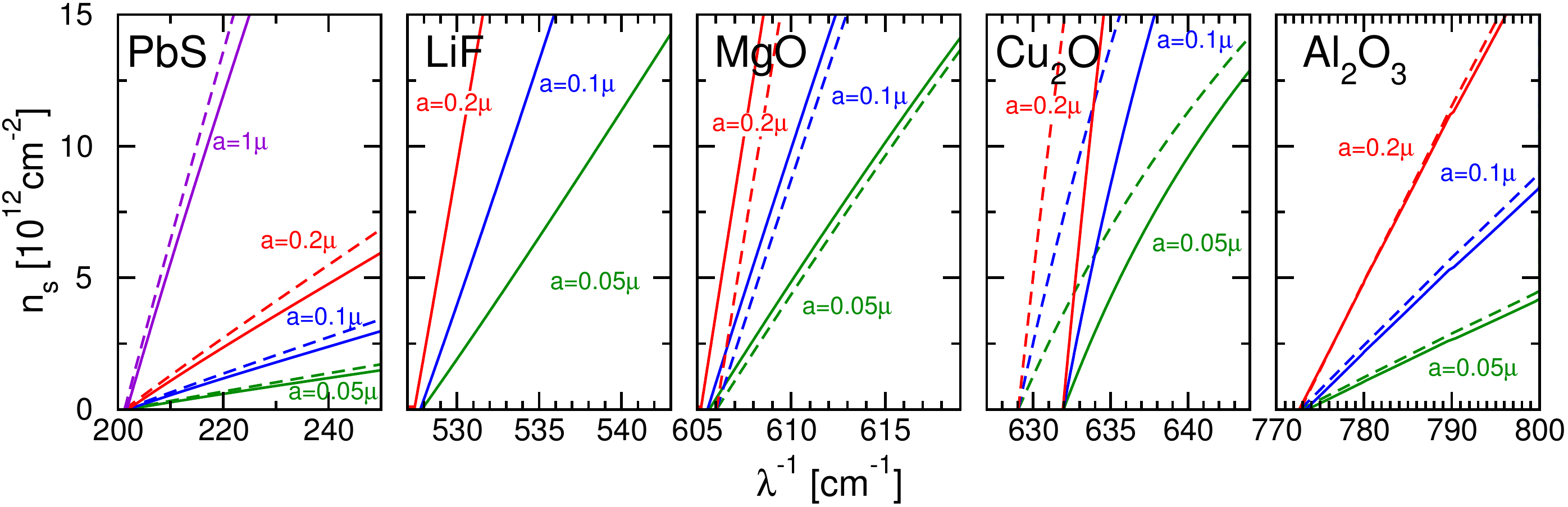}
\caption{(Color online) Position of the extinction resonance depending on the surface charge $n_s$ (or the equivalent bulk charge 
$n_b = 3n_s /a$) for PbS, LiF, MgO,
Cu$_2$O, and Al$_2$O$_3$ particles with different radii $a$. Solid
(dashed) lines are obtained from the Mie contour [Eq. (\ref{rescon_b1})].
}
\label{fig_anres2}
\end{figure*}

A comparison of the resonance shift for MgO and LiF ($\chi<0$) as well as Al$_2$O$_3$, PbS and Cu$_2$O~\cite{matpar} ($\chi>0$) is given by Fig. \ref{fig_anres2}. The shift tends to be larger for bulk ($\chi>0$) than for surface ($\chi<0$) surplus electrons. Cu$_2$O is an example for a dielectric where $\epsilon^{\prime \prime}$ is too large for a well-resolved series of extinction resonances to form. Nevertheless a tail of the lowest resonance for small particles is discernible which is blue-shifted by surplus electrons, albeit by a lesser extent than for Al$_2$O$_3$ or PbS. PbS has a particularly strong resonance shift. Compared to the other materials the TO phonon resonance of PbS is located at a significantly lower frequency where  $\alpha$ is particularly large. Together with the small conduction band effective mass which benefits the electrons' mobility this leads to the larger charge-induced blue-shift.

The blue-shift of the extinction resonance could be used as a charge measurement for particles with $a<1\mu$m. The resonance shift is found for particles with $\chi<0$, e.g. MgO or LiF, and $\chi>0$, e.g. PbS, Cu$_2$O or Al$_2$O$_3$. The most promising candidates are particles made from Al$_2$O$_3$ or PbS. The latter may even allow a measurement for micron-sized particles.

\subsection{Polarization Angles}

So far we have considered charge effects in the extinction efficiency. In the following we will turn to the charge signatures in the polarization of the scattered light. While the extinction (or scattering) efficiency gives only access to the magnitude of the scattering coefficients the polarization of scattered light also reveals the phase of the scattering coefficients. The phase information is particularly useful close to the ordinary and anomalous optical resonances. They occur for $\Re(g_n^{a,b})=0$ where the sign change of $g_n^{a,b}$ leads to a rapid phase change around the resonances. For $\epsilon^{\prime \prime}=0$ the functions $f_n^{a,b}$ and $g_n^{a,b}$ are real in the small particle limit (cf. Eqs. (\ref{fg_a1}) -(\ref{fg_b2})). In this limit $f_n^{a}\sim \rho^{2n+3}$ and $f_n^{b}\sim \rho^{2n+1}$ while $g_n^{a,b}\sim \rho^0$ (for uncharged particles), which entails $g_n^{a,b}>f_n^{a,b}$ except very close to the resonance. As a consequence the phase of the scattering coefficients varies over the resonances by about $\pi$.

For linearly polarized incident light ($\mathbf{E}_i \sim \mathbf{\hat{e}}_x$) the electric field of the reflected light,  
\begin{align}
\mathbf{E}_r\sim & E_0 \frac{e^{-i\omega t +i kr}}{ikr} \sum_{n=1}^\infty \frac{2n+1}{n(n+1)}\nonumber \\ & \times \left[ \left( a_n^r \frac{P_n^1(\cos \theta)}{\sin \theta }+b_n^r \frac{\mathrm{d} P_n^1 (\cos \theta)}{\mathrm{d} \theta} \right) \cos \phi \mathbf{\hat{e}}_\theta \right. \nonumber \\& - \left. \left(a_n^r \frac{\mathrm{d} P_n^1 (\cos \theta)}{\mathrm{d} \theta}+b_n^r \frac{P_n^1 (\cos \theta)}{\sin \theta} \right) \sin \phi \mathbf{\hat{e}}_\phi \right],
\end{align}
 is in general elliptically polarized ($P_n^1(\mu)=\sqrt{1-\mu^2}\mathrm{d}P_n(\mu)/\mathrm{d}\mu$ with $P_n(\mu)$ a Legendre polynomial). Rewriting the reflected electric field as
\begin{align}
\mathbf{E}_r\sim E_0 \frac{e^{-i\omega t +i kr}}{ikr}\left(A_2 e^{i\phi_2} \mathbf{\hat{e}}_\theta +A_3 e^{i\phi_3} \mathbf{\hat{e}}_\phi \right),
\end{align}
where the amplitudes $A_2$, $A_3$ and the phases $\phi_2$, $\phi_3$ are given implicitly by the above equation, the ellipsometric angles are defined by
\begin{align}
\Delta \phi=\phi_2-\phi_3 \quad \text{and} \quad \tan\psi=\frac{A_2}{A_3}.
\end{align}
The angle $\psi$ gives the amplitude ratio and the phase difference $\Delta \phi$ characterizes the opening of the polarization 
ellipse. For $\Delta \phi=0,\pm \pi$ the reflected light is linearly polarized while for $\Delta \phi=\pm \pi/2$ the opening 
of the polarization ellipse is maximal. 

\begin{figure*}[t]
\includegraphics[width=\linewidth]{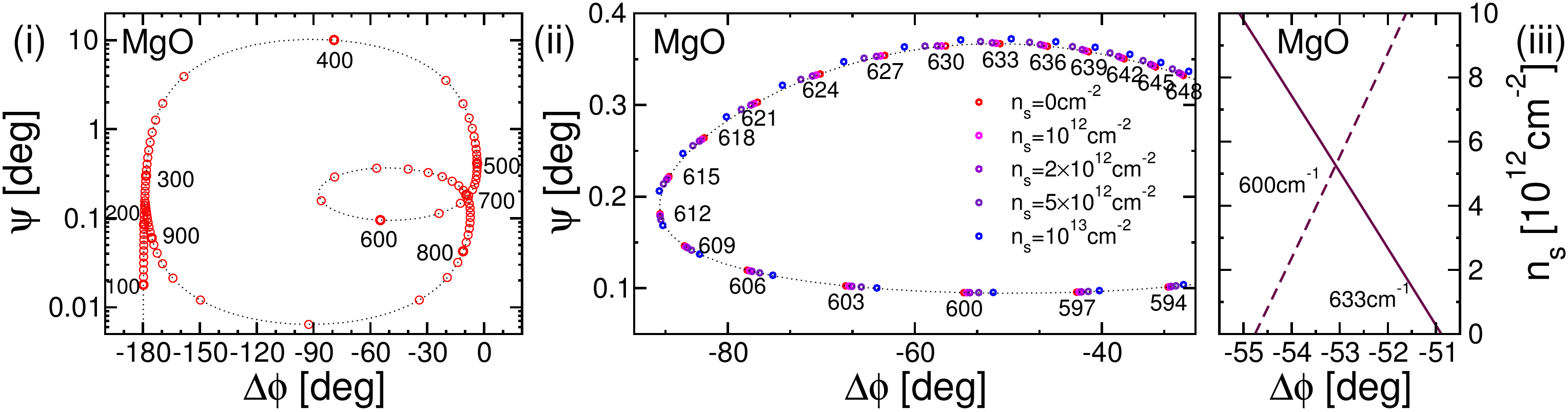}
\includegraphics[width=\linewidth]{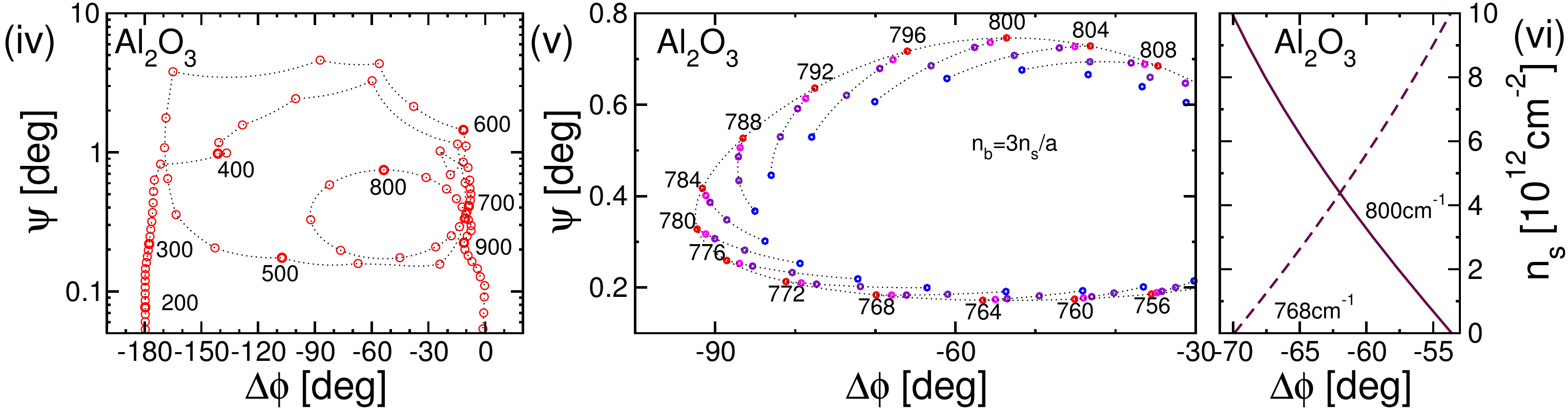}
\caption{(Color online) Ellipsometric angles \(\Psi\) and \(\Delta\phi\) for scattering by an MgO and Al$_2$O$_3$ particle with radius $a=0.5\mu$m in the direction $\theta=\pi/2$ and $\phi=\pi/4$.  (i) (MgO) and (iv) (Al$_2$O$_3$) show \(\Psi\) and \(\Delta\phi\) for $0$ cm$^{-1}<\lambda^{-1}<1000$ cm$^{-1}$ for an uncharged particle. (ii) (MgO) and (v) (Al$_2$O$_3$) magnify the vicinity of the extinction resonance. \(\Psi\) and \(\Delta\phi\) are shifted with increasing surface electron density $n_s$ (or corresponding bulk electron density $n_b=3n_s/a$). The annotated value at the base point gives the wave-number. From there the electron density increases counter-clockwise along the branches. The shift in  \(\Delta\phi\) as a function of \(n_s\) or correspondingly $n_b$ is shown for two representative \(\lambda^{-1}\) in (iii) (MgO) and (vi) (Al$_2$O$_3$).  }
\label{fig_ellips}
\end{figure*}

Note that forward scattered light ($\theta=0$),
\begin{align}
\mathbf{E}_r\sim E_0 \frac{e^{-i\omega t +i kr}}{ikr} \sum_{n=1}^\infty \frac{2n+1}{2}\left(a_n^r+b_n^r \right) \mathbf{\hat{e}}_x,
\end{align}
is linearly polarized. The same applies to backscattered light ($\theta=\pi$) or light that is scattered perpendicularly to the incident wave and in plane or perpendicularly to the direction of polarization of the incident light ($\theta=\pi/2$ and $\phi=0$ or $\phi=\pi/2$).  

An important scattering angle where the scattered light is elliptically polarized is perpendicular to the incident wave and at 45$^\circ$ to the plane of polarization of the incident wave ($\theta=\pi/2$ and $\phi=\pi/4$). This configuration is also used to determine from the Mie signal the particle size of nanodust.\cite{GCK12} 
Figure \ref{fig_ellips} shows the polarization angles $\Delta \phi$ and $\psi$ for this scattering direction for MgO and Al$_2$O$_3$ particles with radius $a=0.5\mu$m. Panels (i) (MgO) and (iv) (Al$_2$O$_3$) give an overview for an uncharged particle. 

In the small particle limit only the scattering coefficients $a_1^r$, $b_1^r$, and $b_2^r$ are relevant. The reflected electric field is given by 
\begin{align}
\mathbf{E}_r\sim E_0 \frac{e^{ikr-i\omega t}}{ikr} \left[ \left(\frac{3}{2\sqrt{2}}a_1^r-\frac{5}{2\sqrt{2}}b_2^r\right)\mathbf{\hat{e}}_\theta-\frac{3}{2\sqrt{2}}b_1^r \mathbf{\hat{e}}_\phi \right].
\end{align}
Figure \ref{fig_ellips} (i) shows a strong variation of $\Delta \phi$ for MgO as a function of $\lambda^{-1}$ which can be related to the variation of the phase of the scattering coefficients. For low frequencies $\lambda^{-1}<300$ cm$^{-1}$ the reflected light is linearly polarized. Close to 400 cm$^{-1}$ the rapid phase variation by $\pi$ of the coefficient $a_1^r$ increases $\Delta \phi$ by about $\pi$. Above $\lambda_{TO}^{-1}$ resonances appear in the coefficients $b_1^r$ and $b_2^r$ for $\epsilon^\prime=-2$  and  $\epsilon^\prime=-3/2$ (for $\rho \ll 1$), respectively.  As these resonances are located very close to each other, the phase shifts by $\pi$ partly cancel and  $\Delta\phi$ acquires and looses a phase of about $-\pi/2$ at around $\lambda^{-1}=600$ cm$^{-1}$. For Al$_2$O$_3$ the variation of $\Delta \phi$ is more complicated because two TO phonon modes dominate $\epsilon$. Nevertheless the interplay of the $b_1$ and the $b_2$ mode above the higher TO phonon resonance leads to  the rapid variation of $\Delta \phi$ from close to 0 to $-\pi/2$ and back to close to 0 near $800$ cm$^{-1}$. 

Surplus charges alter the polarization angles very little except near the rapid opening and closing of the polarization 
ellipse at the anomalous resonances. Here surplus charges lead to a blue shift of the resonances in $b_1^r$ and $b_2^r$. The shift is approximately given by Eq. (\ref{rescon_b1}) for the mode $b_1$ and by $2\epsilon^\prime+2\alpha^\prime+3-6\tau^{\prime \prime}/\rho =0 $  for the mode $b_2$  (in both cases $\rho \ll1$).  The resonance blue-shift translates into a shift of $\Delta \phi$. For a charged particle $\Delta \phi$ acquires and looses $-\pi/2$ as for an uncharged particle but this takes place at higher $\lambda^{-1}$ than for an uncharged particle. This is shown in panels (ii) and (v) of Fig. \ref{fig_ellips}. Panels (iii) and (vi) exemplify it for fixed $\lambda^{-1}$ where $\Delta \phi$ increases or decreases with the particle charge. This shift of $\Delta \phi$ by several degrees should also offer a possibility for a charge measurement.

\section{Conclusions}
\label{sec_summary}

We studied the scattering behavior of a charged dielectric particle with an eye on identifying a strategy for an optical charge measurement. Our focus lay on the  four characteristic regimes of scattering for particles with a strong TO phonon resonance: (i) low-frequency scattering, (ii) ordinary resonances, (iii) anomalous resonances, and (iv) interference and ripple structure.
Surplus charges enter into the scattering coefficients through their phonon-limited surface (for negative electron affinity) or bulk (positive electron affinity) conductivities. 

No significant charge effects are found for the ordinary resonances and the interference and ripple structure. Surplus charges affect however the low-frequency regime and the anomalous optical resonances.

 We have identified three charge-dependent features of light scattering: (i) a charge-induced increase in extinction for low-frequencies, (ii) a blue-shift of the anomalous extinction resonance, and (iii) a rapid variation of one of the two polarization angles at the anomalous extinction resonance. 
At low frequencies energy relaxation is inhibited for uncharged particles as the imaginary part of the dielectric constant is very small. Surplus charges enable energy relaxation on the grain through their electrical conductivity which has a significant real part at low frequencies. This leads to increased absorption at low frequencies.
Above the TO phonon frequency the real part of the dielectric constant is negative which leads to anomalous optical resonances. Surplus charges enter into the resonance condition through the imaginary part of their electrical conductivity. They lead to a resonance blue-shift which is most significant for sub-micron-sized particles. Moreover, at the anomalous resonances the phase of the resonant scattering coefficients varies rapidly. This causes the opening and closing---characterized by the angle $\Delta \phi$---of the polarization ellipse of the reflected light. Surplus charges lead to the rapid variation in $\Delta \phi$ being shifted to higher frequency.

We suggest to use these charge signatures in the Mie signal to measure the particle charge. For plasmonic particles charge-induced resonance shifts have already been detected experimentally for metallic nanorods which were charged by an electrolytic 
solution~\cite{MJG06,NM07} and for an array of nanodiscs exposed to an argon plasma.\cite{LSH12}

In order to detect the charge-sensitive effects of light scattering by dust particles in a dusty plasma would 
require to shine infra-red light through the plasma and to measure light attenuation or the polarization of reflected 
light. The low-frequency increase in extinction or the shift in the polarization angle $\Delta \phi$ could be observed 
with monochromatic light while the resonance shift would require a frequency dependent extinction measurement. This would 
not only allow a determination of the particle charge without knowing any plasma parameters but also of nanodust 
particles~\cite{GCK12,KSB05,BKS09} where traditional techniques cannot be applied at all. 

Eventually suitable particles with a strong charge sensitivity (e.g. Al$_2$O$_3$ or PbS particles) could even be employed as minimally invasive electric plasma probes. The particles would accumulate a charge  depending on the local plasma environment. Performing simultaneously an optical charge measurement and a traditional  force measurement~\cite{KRZ05,TLA00,CJG11} would then allow to infer the local electron density and temperature at the position of the probe particle.

\section*{Acknowledgement}

This work was supported by the Deutsche Forschungsgemeinschaft through SFB-TR 24.

\end{document}